# Picosecond Spin Orbit Torque Switching

Kaushalya Jhuria[1], Julius Hohlfeld[1], Akshay Pattabi[2], Elodie Martin[1], Aldo Ygnacio Arriola Córdova[1,3], Xinping Shi[4], Roberto Lo Conte[2], Sebastien Petit-Watelot[1], Juan Carlos Rojas-Sanchez[1], Gregory Malinowski[1], Stéphane Mangin[1], Aristide Lemaître[5], Michel Hehn[1], Jeffrey Bokor[2,6], Richard B. Wilson[4], Jon Gorchon[1,*]

Reducing energy dissipation while increasing speed in computation and memory is a long-standing challenge for spintronics research[1]. In the last 20 years, femtosecond lasers have emerged as a tool to control the magnetization in specific magnetic materials at the picosecond timescale[2–4]. However, the use of ultra-fast optics in integrated circuits and memories would require a major paradigm shift. An ultrafast *electrical* control of the magnetization is far preferable for integrated systems. Here we demonstrate reliable and deterministic control of the out-of-plane magnetization of a 1 nm-thick Co layer with single 6 ps-wide electrical pulses that induce spin orbit torques on the magnetization. We monitor the ultrafast magnetization dynamics due to the spin orbit torques with sub-picosecond resolution, thus far accessible only by numerical simulations. Due to the short duration of our pulses, we enter a counter-intuitive regime of switching where heat dissipation assists the reversal. Moreover, we estimate a low energy cost to switch the magnetization, below 50 pJ for our micrometer sized device. These experiments show that spintronic phenomena can be exploited on picosecond time-scales for full magnetic control and should launch a new regime of ultrafast spin torque studies and applications.

There is a largely held belief[5,6], that the time it takes for the magnetization of a system to be reversed coherently is limited to half of its natural precessional period (i.e. its ferromagnetic resonance, FMR). However, if a strong enough effective field is induced the Landau-Lifshitz-Gilbert (LLG) equation that governs the magnetization dynamics predicts switching on time-scales shorter than the FMR half-period. It is also commonly claimed[5,6] that ferromagnetic materials have FMR frequencies of a few GHz, limiting switching to a fraction of a nanosecond (ns). But this view neglects the fact that most industrially relevant ferromagnetic systems are thin films with strong perpendicular magnetic anisotropy fields ($H_a$), typically[7] around 1 T (going up to 5 T), which have much higher FMR frequencies and half-periods below $\pi/\gamma H_a \sim 20$ ps ($\gamma$ being the gyromagnetic ratio)[8].

Experimentally, current induced magnetization switching has been demonstrated[9] with current pulses as short as 50 ps using the spin transfer torque (STT), albeit only in in-plane magnetized samples. In STT devices a first magnetic layer polarizes the current, which is then injected into a second softer magnetic layer to impart a torque and switch its magnetization[10]. Generally, the two magnetic layers are spaced by a thin insulating barrier, in order to allow a readout of the magnetic state via the tunneling magneto resistance effect. In practice, switching in STT devices is typically limited to the nanosecond[11] (ns), since shorter current pulses would require increased current densities that would lead to damage of the tunnel barrier[12]. A recent alternative to STT are spin-orbit torques[13,14] (SOT), where the spin polarization is obtained by flowing a current in high spin-orbit materials or interfaces[15]. Because of the geometry of SOT devices, the current does not flow through the barrier, and switching in devices has been demonstrated with pulses as short as 200 ps[12,16].

The limitations of scientific instruments are a major obstacle to switching on time-scales shorter than 200 ps. Commercial electrical current do not have sufficient bandwidth and/or amplitude to switch SOT devices on ps time-scales. However, picosecond-ready CMOS transistors exists since 2007[17] in commercial technology. Therefore, if methods are discovered for using picosecond electrical pulses and SOT to control magnetic order, these methods have the potential to be integrated into mainstream integrated circuits.

To investigate picosecond SOT phenomena, we use an experimental platform that employs photoconductive switches[18] to generate picosecond electrical pulses. Similar devices have been previously used to demonstrate the reversal of the magnetization of a GdFeCo thin-film with a sub-10ps electrical pulse[19]. However, the switching mechanism in GdFeCo relies on ferrimagnetic order and is therefore not generalizable to most magnetic materials. The magnetic moment of GdFeCo toggles back and forth on repeated pulses independent of the current polarity due to the effects of rapid Joule heating[19,20]. For application in devices, one would rather have a mechanism in which the final state depends on the polarity of the current, not on the previous state of the magnetic bit. It is also desirable for the switching mechanism to be compatible with a wide range of materials. Here, we generate and inject 6 ps-wide electrical pulses to trigger SOT dynamics in a prototypical thin Co magnetic film, leading to ultrafast magnetization dynamics and a complete reversal of its magnetic moment.

### Device, setup & quasi-static switching

We deposit a Ta(5nm)/Pt(4)/Co(1)/Cu(1)/Ta(4)/Pt(1) stack (shown in Figure 1.a, with thicknesses in nm) on both glass and GaAs substrates (see methods). The Co layer has perpendicular magnetic anisotropy, as shown by the anomalous Hall effect measurement in Figure 1.c. The bottom Pt and top Ta were chosen for their opposites signs of spin Hall angles, in order to

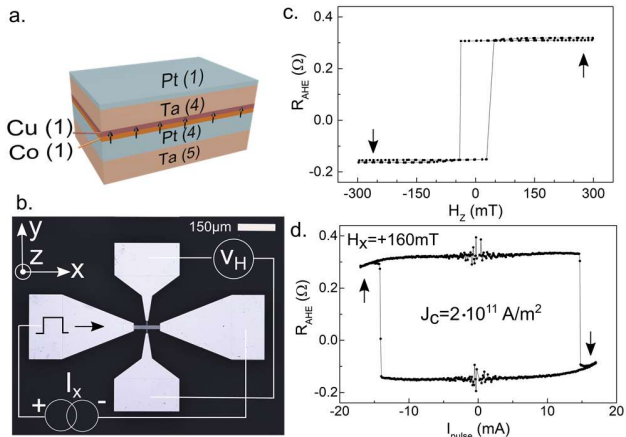

**a.**

Pt (1)
Ta (4)
Cu (1)
Co (1)
Pt (4)
Ta (5)

**b.**

150μm

$V_H$

$I_x$

**c.**

$R_{AHE}$ (Ω)

$H_z$ (mT)

**d.**

$R_{AHE}$ (Ω)

$H_x$=+160mT

$J_C$=2·10$^{11}$ A/m$^2$

$I_{pulse}$ (mA)

**Figure 1: Sample and switching behavior via field and current.** a) Magnetic sample stack b) Patterned Hall bar using the magnetic stack and gold contact pads. The schematic shows the electrical connections used for Anomalous Hall resistance ($R_{AHE} = V_H/I$) detection as a function of c) out-of-plane magnetic field ($H_z$) and d) 100 μs current pulses under a 160mT in-plane field ($H_x$).

enhance the torques on the Co layer[21] (see methods for sample details). We fabricate the Hall structures shown in Figure 1.b, and perform field driven (Figure 1.c) and current driven (Figure 1.d) magnetic hysteresis loops by monitoring the Hall resistance. For the current driven case we use 100 μs-long current pulses while a uniform and constant in-plane symmetry-breaking magnetic field $H_x$ is applied. We find that the critical current density $J_c$ for switching the magnetization is equal to ~$2 \cdot 10^{11}$A/m$^2$ for a 160 mT field, and is inversely proportional to the in-plane field as expected from SOT-driven switching[13,22,23] (see supp. mat.). The sample switches to $-M_z$ ($+M_z$) when the in-plane magnetic field and the charge current are parallel (antiparallel), in agreement with a SOT rising from the combination of the spin Hall effect from both heavy metals[24,25].

We use low-temperature GaAs (LT-GaAs) photoconductive switches to generate picosecond pulses of high intensity[18] (shown in Figure 2, see methods for sample fabrication and pulse generation details). The bias voltage $\Delta V$ allows us to select the amplitude and polarity of the current. In order to generate high current pulses we excite the switches with an amplified 5 kHz repetition rate laser system with 30 fs laser pulses centered at

800 nm. The use of this laser system with our photoswitches results in 6 ps duration ($\tau_p$) high intensity current pulses (shown in the graph of Figure 2). We measure the pulse duration with a Teraspike® free-standing electric field detector (see methods). After excitation, the electrical pulses propagate (see Figure 2) on Au coplanar waveguides and are focused into the magnetic structure by an impedance matched taper.

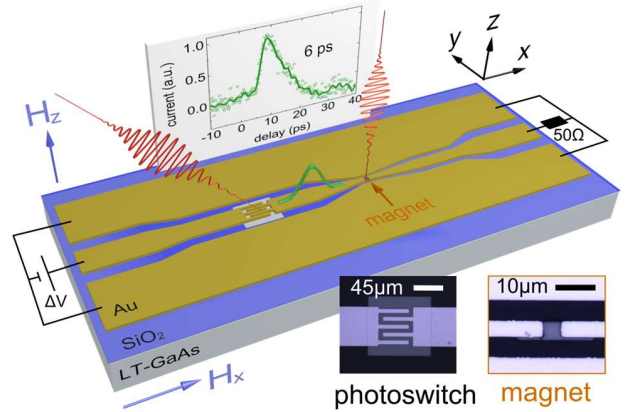

**Figure 2: Setup for generation of picosecond electrical pulses and magneto-optical detection.** The optical pump excites the photoconductive switch in order to generate ~6 ps duration electrical pulses, that are guided and focused by a coplanar waveguide into the magnetic stack, resulting in ultrafast spin orbit torques. The sampled picosecond current pulse is shown on the back of the figure. The solid green line is a guide for the eyes.

## Switching with a single 6ps –wide pulse

Figure 3 shows polar magneto optical Kerr effect (MOKE) micrographs of the initial configuration and final state after a single 6ps electrical pulse, for various configurations. In each of the four quadrants of Figure 3 we test combinations of current $I$ and in-plane magnetic field $H_x$ directions. Again, we observe that parallel (antiparallel) current pulses and field result in $-Mz$ ($+Mz$), just as expected by the symmetries of the SOTs in the prepared stack. Moreover, we observe that the final state is independent of the initial magnetic state of our stack. We injected up to 10 successive pulses of the same polarity and saw no difference in the final state. We successfully repeated the experiment (initial saturation + single shot) at the switching voltage for $n =35$ times. We can thus estimate a >91% switching probability with a 95% confidence interval (as per the "rule of three", $P > 1 - 3/n$). As soon as the bias voltage $\Delta V$ is decreased below 40V, no more reversal is observed. When the in-plane field is reduced below ($H_x \sim 120$ mT) no reversal is observed, likely requiring higher current densities, as also observed by Garello et al.[12]. In this work, we did not explore higher current densities in order to avoid the risk of degradation or permanent damage of the photoswitch. To test the device endurance, we subjected the device to electrical pulses at switching conditions ($\Delta V \sim 40V$), a repetition rate of 5kHz, for 5 hours ($> 10^8$ pulses). After the endurance testing we noticed no degradation or change to the electrical and magnetic properties.

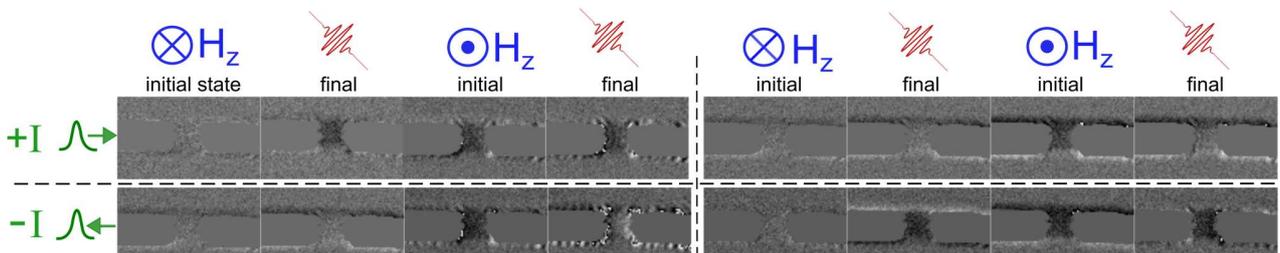

**Figure 3: MOKE micrographs of single 6 ps electrical pulses switching the magnetization via SOT.** The four quadrants show 2 before and 2 after-pulse images under different in-plane field and current directions. The inversion of the final state with current or in-plane field is a clear signature of SOT switching. Bias voltages used for switching were slightly above the critical threshold ($\Delta V \sim 40$ V). Light (dark) grey indicates magnetization down (up).

**Time-resolved dynamics due to a 3.7ps-wide pulse**

To measure the ultrafast magnetic response to the current pulses, we perform time-resolved MOKE measurements under various configurations of current polarity and magnetic fields. We were unable to obtain time-resolved switching dynamics due to technical limitations (see supp. mat.). We therefore perform low intensity time-resolved studies with an 80 MHz oscillator laser with $\sim$250 fs-duration pulses centered at 780 nm (see methods). With this system we obtain 3.7 ps-duration electrical pulses (inset in .a). In these experiments, we monitor the change in the out-of-plane component of the magnetization ($\Delta M_z$) via polar-MOKE with a time-delayed ($\Delta t$) probe pulse. The typical magnetic response to the pulses is shown in .a.

We first focus our attention on the dynamics under zero in-plane field (black symbols in .a). The response of the Co magnetization to the current is instantaneous (see methods for time-delay $\Delta t$ calibration). Both for $+M_z$ and $-M_z$, the out-of-plane magnetization abruptly decreases, and slowly recovers. The decrease of $|M_z|$ is due to two mechanisms. First, the SOT pulls the magnetization towards the plane. Second, the picosecond charge current induces Joule heating which leads to an ultrafast loss of magnetic order[20,26], commonly known as ultrafast demagnetization[20,26]. The latter mechanism, and the subsequent slow cooling by heat diffusion, explain the slow recovery at long time delays (350 ps, see black curve in .a and .b).

We now focus on the dynamics under in-plane magnetic fields (blue and red symbols in .a). Unlike in the zero field case (black symbols in .a), the magnetic moment is not initially oriented along $z$. In the presence of an in-plane field pointing along $x$, the moment at negative time-delays is tilted in the $xz$ plane, to be parallel with the effective field $H_{eff}(\Delta t < 0)$, see .c. The in-plane field breaks the symmetry of the system and, together with the injected spin polarization $\sigma_y$, determines the sign for the observed coherent precession. A parallel (antiparallel) in-plane field and current causes the moment to process towards (away from) $-z$, causing $\Delta M_z$ to decrease (increase) on a 10 ps time-scale. This occurs regardless of the initial up ($+M_z$),or down ($-M_z$) state, as expected from SOT and in agreement with the result of our quasi-static SOT switching experiments from Figure 1.d.

The precessions in .a are offset by the heating induced demagnetization (reduction of $|M_z|$). In addition to reducing the magnetization, ultrafast Joule heating reduces the magnetic anisotropy. We also depict this effect in .e. As temperature rises, the anisotropy field $H_a(T)$ drops and, under a constant external field, the angle (and amplitude) of the effective field $H_{eff}(\Delta t > 4ps)$ changes. The moment experiences a torque $\tau_{H_{eff}}$ due to the change in angle of $H_{eff}$. The torque $\tau_{H_{eff}}$ is enhanced by the SOT $\tau_{DL}$ which pulls the moment even further from $H_{eff}$. The torque $\tau_{H_{eff}}$, which we call the thermal anisotropy torque, is commonly used in ultrafast pump-probe FMR experiments as a way to trigger oscillatory dynamics[27,28]. Surprisingly, the thermal anisotropy torque alone can lead to a complete switching of the magnetization of a ferromagnet, as it was recently demonstrated using femtosecond optical pulses[29]. In our experiments, the thermal anisotropy torque can assist the SOT in the switching of the magnetization.

**Model**

To understand the ultrafast dynamics, we use a simple LLG macrospin model. The model includes SOTs and ultrafast Joule heating (for details see suppl. mat.). The model assumes simplistic temperature dependence laws for the anisotropy and magnetization to calculate the thermal anisotropy torques. We set the spin Hall angle to 0.3 from Ref [30], $M_s$ from vibrating sample magnetometry (VSM) measurements. We set the damping and anisotropy at room temperature using optically excited time-resolved MOKE measurements[27]. We also had to include a number of electrical reflections of the current pulses from the end of the transmission lines which affect the dynamics. The resulting best fits are shown in .b. The quality of the fit is remarkable for such a simple model.

We now describe the model predictions of the dynamics in response to current pulses in the presence of an in-plane field. The model predictions are shown in .b-e: The magnetization $\boldsymbol{m}$ is initially in its equilibrium position, along $H_{eff}(\Delta t < 0)$, as in .c. As soon as the current pulse arrives, a damping-like SOT[31] $\tau_{DL} \sim \boldsymbol{m} \times \boldsymbol{m} \times \sigma_y$ brings the magnetization towards the $y$ axis regardless of the injected spins $\sigma_y$ (initial drop in $\Delta M_z$ on blue & red curves in .b), as depicted in .d. At the same time heating changes the effective field by decreasing $H_a$. As $\boldsymbol{m}$ is torqued away from its initial position, precession around the evolving effective field begins (the so called thermal anisotropy torque, shown in .e). The two current polarities will lead to a $180°$ phase difference in the precessional dynamics, resulting in opposite $\Delta M_z$ (red and blue trajectories in .b-c). It is interesting to note that a field-like SOT[31] $\tau_{FL} \sim \boldsymbol{m} \times \sigma_y$ cannot reproduce the initial drop in $\Delta M_z$ that leads to a kink close to $\Delta t = 0$ on the blue curve. A damping-like dominant torque agrees well with reports on similar structures[30].

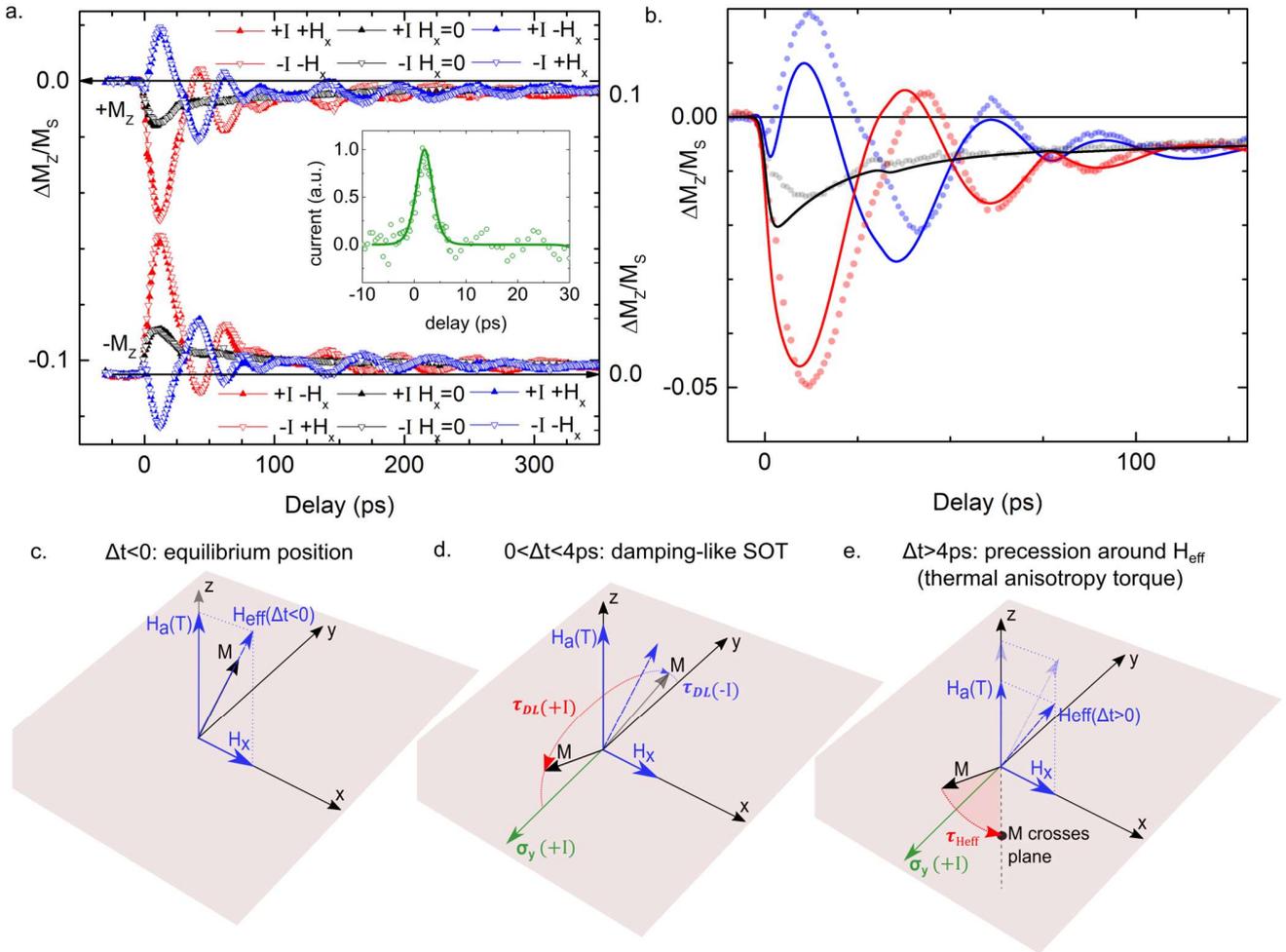

**Figure 4: Time-resolved MOKE response due to a 3.7ps electrical pulse and macrospin simulation.** a) The dynamics include spin orbit torques and thermal effects (demagnetization). The phase and sign of the torques is dependent on the in-plane field ($H_x$) and current direction, as expected from SOT. Without the symmetry-breaking in-plane field the oscillations disappear. The solid green line is a fit of current trace with a sech² function of 3.7ps (FWHM). Experiments were done after $\pm M_z$ saturation of the sample, under $H_z = 0$ mT , $H_x = \pm 160$ mT and a bias voltage of $\Delta V \sim 30$ V. b) Macrospin simulation (lines) including ultrafast demagnetization and SOTs on top of the $+M_z$ data (circles). All data is normalized by the saturation magnetization $M_s$ at room temperature. c-e) Schematic of the magnetization dynamics induced by the ps current pulse as predicted by the model. Description in the text.

Even though the LLG model describes the overall dynamics, its predictions don't agree with certain features of the data. In particular, the model does not match the dynamics of the black trace at $H_x$=0 between 3 and 10 ps. A possible explanation is inhomogeneous broadening. In the time-resolved MOKE measurements of damping and anisotropy we observed that the samples have large inhomogeneous broadening in the effective damping, which is consistent with prior pump/probe studies of dynamics of perpendicularly magnetized films[28]. In addition to inhomogeneities in anisotropy, it is possible that there are spatial inhomogeneities in the excitation, either due to the (spin) current distribution or hot spots. More experimental and theoretical work will be required to better understand these discrepancies.

Simulations performed with the same parameters at higher currents are presented in the suppl. materials, and suggest the thermal anisotropy torque plays an important role for switching. Including the effect of thermal anisotropy torques in the LLG simulations reduces the energy required for switching by a factor of two. A switching (crossing $M_z = 0$) as fast as 16ps is predicted, with a recovery of 80% of the magnetization reached after 50ps. Finally, we highlight that just as in Ref [29], our simulations can lead to switching even without spin-torques (setting spin Hall angles to zero), only due to the thermal anisotropy torque (see supplementary materials).

## Energy estimation

We can set an upper limit to the energy dissipated at the load by estimating the total initial energy stored in the photoswitch capacitor (see methods), which is the energy that drives both the THz current pulse in the lines and the losses to THz radiation[32~35].

The estimated maximum energy consumption at the magnet (neglecting all losses) is of $50$ pJ. This in turn corresponds to a maximum critical current density of $J_c \sim 6 \cdot 10^{12}$ A/m$^2$. Therefore, the energy requirements in this ultrafast SOT regime compare extremely favorably with state of the art, ns and sub-ns switching SOT results[16,36] and other types of memory[36], even though we are using a non-optimized stack with relatively large μm dimensions.

## Discussion of possible mechanisms

Switching via a spin (orbit) torque on such a fast timescale brings up interesting questions about the mechanisms involved. According to our fits, the low current dynamics are almost fully coherent. However, as we approach the switching currents, more complex mechanisms such as magnetic domain nucleation and magnetic domain wall propagation might play a role. We will discuss some of these mechanisms in the following paragraphs. In particular, we will distinguish stochastic nucleation due to thermal noise and deterministic nucleation due to geometry, defects or spatial inhomogeneities in the sample properties.

For long pulses it is well known that thermal activation plays an important role in the nucleation of reversed magnetic domains, which followed by domain wall motion allows for full switching[12]. Thermal activation, and in particular the attempt frequency, is generally thought to be correlated to the FMR frequency[37]. In fact, the noise spectrum of a ferromagnet under a DC current has been experimentally found to peak at the FMR[38,39]. Therefore, when using current pulses wider than the FMR period ($\sim 40$ ps, for $H_a \sim 1$ T), some degree of stochasticity should be expected. It is thus surprising that many works assume thermal activation should be negligible in the sub-ns regime[12,16,40–42]. Stochasticity in the context of sub-ns SOT switching has only been very recently reported[41,42]. If we now consider our experiments with 6ps pulses, well below the FMR period of our samples, stochasticity becomes more questionable. An important point to keep in mind is that with ps pulses the transient temperature rise (and thus thermal noise amplitude) in the film can be much larger than with ns pulses. In fact, thermal noise terms are successfully included in models for optically induced ps switching dynamics[2,43–45], but ultimately have a macroscopic deterministic role. Experimentally, a few-nm sized nucleation points have been observed at the ps timescale[43,44], but are explained as a magnon localization[44] rather than classical domain nucleation (we remind that domain wall widths are typically ~5-10nm in equilibrium, and that a domain should be larger than two walls). At those length-scales, the dynamics resemble those of a temperature dependent macrospin model[45]. In summary, the question of the role of thermal fluctuations at ultra-fast timescales is far from resolved, but we believe it is unlikely that stochastic nucleation of classical domains is taking place in our experiments.

If we then consider the switching to be purely deterministic, a few different scenarios are possible. We can first consider an edge-to-edge single domain wall sweeping scenario, assuming a deterministic nucleation at one sample edge. Recently, it was experimentally shown[46] that even in large (4μm long) devices,

SOT reversal with 0.2ns pulses could be driven by a single domain wall motion. Given our ultra-short pulses, this scenario would imply unphysical domain wall velocities[47,48] of $10^5$ m/s, too large for the estimated current densities.

A second possible scenario would be one considering multiple domain nucleations at specific sample sites, due to spatial inhomogeneities in material properties. Based on our estimated peak current density ($J_c \sim 6 \cdot 10^{12}$ A/m$^2$), if we extrapolate the expected domain wall velocity based on results of the fastest measured velocities in ferromagnets[47] we obtain ~1000 m/s. A Gaussian-like current pulse of 6 ps would allow for less than ~6nm of displacement, which is comparable to the domain wall width[48]. Therefore, under this scenario switching would be driven by nucleation. Nucleation would then need to cover at least 50% of the area, in order to determine the final state. At longer timescales, these domains could merge leading to a full switching[41]. We note that we did observe some domain formation observed in a different device, at currents close to the switching threshold, possibly correlated to the current distribution (see suppl. mat.). In order to check for a signature of non-coherent switching at ultrafast timescales, we performed low intensity time-resolved dynamics at various positions on the magnet with our ~1.5 μm diameter probe and found negligible differences (see suppl. mat.). This measured homogeneity could mean nucleated domains are too small and numerous, or just that spatial inhomogeneities in the dynamics are below the sensitivity at such low intensities.

A final alternative scenario would be to consider some form of heat-assisted magnetic recording due to Joule heating. In such scenario the anisotropy barrier and/or magnetic order would be completely reduced by heating up to the Curie temperature $T_c$ and any tiny torque could then determine the switching direction. To test for this scenario, we performed experiments where the in-plane field was replaced by an out-of-plane field, just below the coercive field and favoring a reversal (see suppl. mat.). No reversal was observed after injection of a single current pulse at the critical current, which allows us to discard the HAMR scenario.

Therefore, we believe that, depending on the homogeneity of the sample properties, the switching in our samples could be either nearly-coherent or governed by multiple nucleation events that happen within the current pulse and cover at least 50% of the area. Tracking the dynamics during the reversal could help in distinguishing between these mechanisms.

## Conclusions

In conclusion, we have demonstrated spin orbit torque switching of a thin Co film with a single $6$ ps electrical pulse. We show that picosecond-duration electrical pulses can inject spin into a magnet at ultrafast timescales. We can then probe the generated magnetization torques with picosecond resolution. All our experiments are in agreement with the symmetries expected from SOT. Macrospin simulations can accurately predict the observed dynamics, showing a damping-like torque dominated effect. Finally, we have shown that the reversal process is extremely energy efficient. Future work will include tracking different components of the magnetization via different magneto-optical effects, in order to spatially reconstruct the time-dependent spin torque dynamics. We believe our approach will trigger new interest in ultrafast electrical studies of spin torque dynamics, opening the door for the possible observation of elusive phenomena such as inertial dynamics in ferromagnetic materials[49–51] and offer a new way of triggering resonant dynamics in antiferromagnetic materials[52,53].

## References

1.  Åkerman, J. Toward a universal memory. *Science.* **308**, 508–510 (2005).
2.  Radu, I. *et al.* Transient ferromagnetic-like state mediating ultrafast reversal of antiferromagnetically coupled spins. *Nature* **472**, 205–208 (2011).
3.  Stupakiewicz, A., Szerenos, K., Afanasiev, D., Kirilyuk, A. & Kimel, A. V. Ultrafast nonthermal photo-magnetic recording in a transparent medium. *Nature* **542**, 71–74 (2017).
4.  Kirilyuk, A., Kimel, A. V & Rasing, T. Ultrafast optical manipulation of magnetic order. *Rev. Mod. Phys.* **82**, (2010).
5.  Olejník, K. *et al.* Terahertz electrical writing speed in an antiferromagnetic memory. *Sci. Adv.* **4**, 1–9 (2018).
6.  Manchon, A. *et al.* Current-induced spin-orbit torques in ferromagnetic and antiferromagnetic systems. *Rev. Mod. Phys.* **91**, 035004 (2019).
7.  Dieny, B. & Chshiev, M. Perpendicular magnetic anisotropy at transition metal/oxide interfaces and applications. *Rev. Mod. Phys.* **89**, (2017).
8.  Kent, a. D., Özyilmaz, B. & Del Barco, E. Spin-transfer-induced precessional magnetization reversal. *Appl. Phys. Lett.* **84**, 3897–3899 (2004).
9.  Lee, O. J., Ralph, D. C. & Buhrman, R. A. Spin-torque-driven ballistic precessional switching with 50 ps impulses. *Appl. Phys. Lett.* **99**, 10–13 (2011).
10. Stiles, M. & Zangwill, A. Anatomy of spin-transfer torque. *Phys. Rev. B* **66**, 014407 (2002).
11. Sato, H. *et al.* 14ns write speed 128Mb density Embedded STT-MRAM with endurance>10 10 and 10yrs retention@85°C using wound low damage MTJ integration process. *Int. Electron Devices Meet. IEDM* **2018-Decem**, 27.2.1-27.2.4 (2019).
12. Garello, K. *et al.* Ultrafast magnetization switching by spin-orbit torques. *Appl. Phys. Lett.* **105**, 1–12 (2014).
13. Miron, I. M. *et al.* Perpendicular switching of a single ferromagnetic layer induced by in-plane current injection. *Nature* **476**, 189–193 (2011).
14. Liu, L. *et al.* Spin-Torque Switching with the Giant Spin Hall Effect of Tantalum. *Science.* **336**, 555–558 (2012).
15. Sinova, J., Valenzuela, S. O., Wunderlich, J., Back, C. H. & Jungwirth, T. Spin Hall effects. *Rev. Mod. Phys.* **87**, 1213–1260 (2015).
16. Garello, K. *et al.* SOT-MRAM 300MM Integration for Low Power and Ultrafast Embedded Memories. in *2018 IEEE Symposium on VLSI Circuits* **54**, 81–82 (IEEE, 2018).
17. Mistry, K. *et al.* A 45nm Logic Technology with High-k+ Metal Gate Transistors, Strained Silicon, 9 Cu Interconnect Layers, 193nm Dry Patterning, and 100% Pb-free Packaging. *Proc. IEDM* 247–250 (2007).
18. Auston, D. H., Johnson, A. M., Smith, P. R. & Bean, J. C. Picosecond optoelectronic detection, sampling, and correlation measurements in amorphous semiconductors. *Appl. Phys. Lett.* **37**, 371 (2008).
19. Yang, Y. *et al.* Ultrafast magnetization reversal by picosecond electrical pulses. *Sci. Adv.* **3**, e1603117 (2017).
20. Wilson, R. B. *et al.* Electric current induced ultrafast demagnetization. *Phys. Rev. B* **96**, 045105 (2017).
21. Woo, S. *et al.* Enhanced spin-orbit torques in Pt / Co / Ta heterostructures Enhanced spin-orbit torques in Pt / Co / Ta heterostructures. *Appl. Phys. Lett.* **105**, 212404 (2014).
22. Cubukcu, M. *et al.* Spin-orbit torque magnetization switching of a three-terminal perpendicular magnetic tunnel junction. *Appl. Phys. Lett.* **104**, (2014).
23. Lo Conte, R. *et al.* Spin-orbit torque-driven magnetization switching and thermal effects studied in Ta\CoFeB\MgO nanowires. *Appl. Phys. Lett.* **105**, 122404 (2014).
24. Liu, L., Lee, O., Gudmundsen, T., Ralph, D. & Buhrman, R. A. Current-Induced Switching of Perpendicularly Magnetized Magnetic Layers Using Spin Torque from the Spin Hall Effect. *Phys. Rev. Lett.* **109**, 096602 (2012).
25. Baumgartner, M. *et al.* Spatially and time-resolved magnetization dynamics driven by spin-orbit torques. *Nat. Nanotechnol.* **12**, 980–986 (2017).
26. Beaurepaire, E., Merle, J.-C., Daunois, A. & Bigot, J.-Y. Ultrafast spin dynamics in ferromagnetic nickel. *Phys. Rev. Lett.* **76**, 4250–4253 (1996).
27. Koopmans, B., Ruigrok, J., Longa, F. & de Jonge, W. Unifying Ultrafast Magnetization Dynamics. *Phys. Rev. Lett.* **95**, 267207 (2005).
28. Lattery, D. M. *et al.* Quantitative analysis and optimization of magnetization precession initiated by ultrafast optical pulses. *Appl. Phys. Lett.* **113**, (2018).
29. Davies, C. S. *et al.* Anomalously Damped Heat-Assisted Route for Precessional Magnetization Reversal in an Iron Garnet. *Phys. Rev. Lett.* **122**, 27202 (2019).
30. Woo, S., Mann, M., Tan, A. J., Caretta, L. & Beach, G. S. D. Enhanced spin-orbit torques in Pt/Co/Ta heterostructures. *Appl. Phys. Lett.* **105**, 212404 (2014).
31. Garello, K. *et al.* Symmetry and magnitude of spin-orbit torques in ferromagnetic heterostructures. *Nat. Nanotechnol.* **8**, 587–93 (2013).
32. Hoffmann, M. C. & Fülöp, J. A. Intense ultrashort terahertz pulses: Generation and applications. *J. Phys. D. Appl. Phys.* **44**, 083001 (2011).
33. Bonvalet, A. & Joffre, M. Terahertz Femtosecond Pulses. in *Femtosecond Laser Pulses* (ed. Rullière, C.) 309–333 (Springer, 2005).
34. Gregory, I. S. *et al.* Optimization of photomixers and

antennas for continuous-wave terahertz emission. *IEEE J. Quantum Electron.* **41**, 717–728 (2005).

35. Burford, N. M. & El-Shenawee, M. O. Review of terahertz photoconductive antenna technology. *Opt. Eng.* **56**, 010901 (2017).

36. Wong, H.-S. P. *et al.* Standford Memory Trends. Available at: https://nano.stanford.edu/stanford-memory-trends. (Accessed: 11th November 2019)

37. Brown, W. F. Thermal Fluctuations of a Single-Domain Particle. *Phys. Rev.* **130**, 1677–1686 (1963).

38. El Hadri, M. S. *et al.* Two types of all-optical magnetization switching mechanisms using femtosecond laser pulses. *Phys. Rev. B* **94**, 064412 (2016).

39. Guan, Y. *et al.* Thermal-magnetic noise measurement of spin-torque effects on ferromagnetic resonance in MgO-based magnetic tunnel junctions. *Appl. Phys. Lett.* **95**, (2009).

40. Cubukcu, M. *et al.* Ultra-Fast Perpendicular Spin–Orbit Torque MRAM. *IEEE Trans. Magn.* **54**, 1–4 (2018).

41. Decker, M. M. *et al.* Time Resolved Measurements of the Switching Trajectory of Pt/Co Elements Induced by Spin-Orbit Torques. *Phys. Rev. Lett.* **118**, 257201 (2017).

42. Grimaldi, E. *et al.* Single-shot dynamics of spin–orbit torque and spin transfer torque switching in three-terminal magnetic tunnel junctions. *Nat. Nanotechnol.* **15**, 111–117 (2020).

43. Graves, C. E. *et al.* Nanoscale spin reversal by non-local angular momentum transfer following ultrafast laser excitation in ferrimagnetic GdFeCo. *Nat. Mater.* **12**, 293–298 (2013).

44. Iacocca, E. *et al.* Spin-current-mediated rapid magnon localisation and coalescence after ultrafast optical pumping of ferrimagnetic alloys. *Nat. Commun.* **10**, (2019).

45. Atxitia, U., Nieves, P. & Chubykalo-Fesenko, O. Landau-Lifshitz-Bloch equation for ferrimagnetic materials. *Phys. Rev. B* **86**, 104414 (2012).

46. Cai, K. *et al.* Ultrafast and energy-efficient spin–orbit torque switching in compensated ferrimagnets. *Nat. Electron.* **3**, 37–42 (2020).

47. Miron, I. M. *et al.* Fast current-induced domain-wall motion controlled by the Rashba effect. *Nat. Mater.* **10**, 419–23 (2011).

48. Thiaville, A., Rohart, S., Jué, É., Cros, V. & Fert, A. Dynamics of Dzyaloshinskii domain walls in ultrathin magnetic films. *Epl* **100**, (2012).

49. Kikuchi, T. & Tatara, G. Spin dynamics with inertia in metallic ferromagnets. *Phys. Rev. B* **92**, 1–15 (2015).

50. Wegrowe, J.-E. & Ciornei, M.-C. Magnetization dynamics, gyromagnetic relation, and inertial effects. *Am. J. Phys.* **80**, 607–611 (2012).

51. Neeraj, K. *et al.* Experimental evidence of inertial dynamics in ferromagnets. *arXiv:1910.11284* 1–10 (2019).

52. Němec, P., Fiebig, M., Kampfrath, T. & Kimel, A. V. Antiferromagnetic opto-spintronics. *Nat. Phys.* **14**, 229–241 (2018).

53. Baltz, V. *et al.* Antiferromagnetic spintronics. *Rev. Mod. Phys.* **90**, 15005 (2018).

54. Woo, S. *et al.* Observation of room-temperature magnetic skyrmions and their current-driven dynamics in ultrathin metallic ferromagnets. *Nat. Mater.* **15**, 501–506 (2016).

55. Nan, T. *et al.* Comparison of spin-orbit torques and spin pumping across NiFe/Pt and NiFe/Cu/Pt interfaces. *Phys. Rev. B* **91**, (2015).

56. Ostwal, V., Penumatcha, A., Hung, Y. M., Kent, A. D. & Appenzeller, J. Spin-orbit torque based magnetization switching in Pt/Cu/[Co/Ni]5 multilayer structures. *J. Appl. Phys.* **122**, (2017).

57. Pham, T. H. *et al.* Thermal Contribution to the Spin-Orbit Torque in Metallic-Ferrimagnetic Systems. *Phys. Rev. Appl.* **9**, 1–9 (2018).

58. Gupta, K. C., Ramesh, G., Bahl, I. & Bhartia, P. *Microstrip Lines and Slotlines*. (1996).

## Acknowledgments

We would like to thank Emmanuel Vatoux, Tom Ferté and both Laurent Badie and Gwladys Lengaigne for the electromagnet construction, VSM measurements and sample preparation, respectively. We would like to specially thank Yang Yang and Charles-Henri Lambert for their help in the first SOT switching trials years ago. This work was supported by the impact project LUE-N4S part of the French PIA project "Lorraine Université d'Excellence", reference ANR-15IDEX-04-LUE and the "FEDER-FSE Lorraine et Massif Vosges 2014-2020", a European Union Program. This work was also partly supported by the French RENATECH network. R.L.C. and J.B. gratefully acknowledge support from the National Science Foundation (NSF) through the Cooperative Agreement Award EEC-1160504 for Solicitation NSF 11-537 (TANMS). A.P. and J.B. also gratefully acknowledge support from the NSF Center for Energy Efficient Electronics (E3S). Work by X. S. and R. W. was supported by the U.S. Army Research Laboratory and the U.S. Army Research Office under contract/grant number W911NF-18-1-0364. J. B. also gratefully acknowledges support by ASCENT (one of the SRC/DARPA supported centers within the JUMP initiative). Preliminary experiments in this work were supported by U.S. Department of Energy, Office of Science, Office of Basic Energy Sciences, Materials Sciences and Engineering Division under Contract No. DE-AC02-05-CH11231 within the Nonequilibrium Magnetic Materials Program (MSMAG).

## Author contributions

J.G. designed the experiments with inputs from R.B.W, J.B, S.P-W. & J.H. A.L. grew the LT-GaAs substrates. M.H. optimized and grew the samples by sputtering. K. fabricated the devices. K., J.H. and J.G., performed the ultrafast SOT experiments and characterized the picosecond pulses. E.M., A.Y.A.C. performed the anomalous Hall measurements and 100μs SOT switching experiments under the supervision of J.C.R.S. and S.P-W. who designed the mask and set up this experiment. R.B.W. built the numerical model and performed the simulations with inputs from J.G. R.B.W. performed optical time-resolved experiments to determine the damping and anisotropy of the samples. J.G. analyzed the experimental data with help from K., R.B.W., R.L.C., J.H. and S-P.W. J.G. wrote the manuscript with input from all authors.

## Competing financial interests

The authors declare no competing financial interests.

## Methods

### Samples

The LT GaAs substrate was obtained by first depositing, in a molecular beam epitaxy chamber, at high temperature (550°C) a 300 nm thick GaAl$_{0.8}$As buffer followed by a 5nm thick GaAs layer on a semi-insulating GaAs (100) substrate. Then the LT GaAs layer (1 μm thick) was deposited at 260°C with a As/Ga beam equivalent pressure ratio of 50.

The magnetic stacks were grown by DC magnetron sputtering in an AJA system. The Ta(5nm)/Pt(4) buffer layer ensures a well-defined (111) texture for the growth of Co(1) and guarantees an interface anisotropy that promotes PMA for Co(1). The Cu(1)/Ta(4) bilayer, capped by Pt(1) to prevent Ta oxidation, has been added to preserve PMA and enhance the torques on the Co layer since Pt and Ta have spin Hall angles with opposite sign. First trials with Pt/Co/Ta stacks resulted in non-square magnetic hysteresis curves with small remanence indicating a possible large effective Dylazhosinkii-Moriya interaction (DMI) as in Ref.[54]. In order to obtain two well-defined remanent states at zero field we inserted the Cu layer to reduce the DMI at the top Co interface. Due to the long spin-diffusion length of Cu, spin currents generated in the Ta are expected to contribute to the SOT[55,56]. The choice of the stack was also determined by the necessity of having a top metallic layer (Cu/Ta + Pt capping) in order to get a good electrical contact with the transmission lines shown in Figure 2. The $T_c$ of the sample is estimated as ~800 K due to previous experience with extremely similar samples grown and characterized in the group over the years.

The sample was fabricated using a 3 step-based UV-lithography technique where the SiO$_2$ layer, magnetic load, and transmission lines were patterned at each step. The SiO$_2$ layer allows for a good insulation of the transmission lines from the substrate in order to suppress leakage currents. A single recipe was used to perform lithography and lift-off for all three steps. An Ac 450 (Alliance Concept) sputtering system was used to deposit 100 nm of SiO2 in the presence of 20sccm Ar and O2 flow at a base pressure of 6.1×10-3 mbar. E-beam evaporation was used to deposit Ti(20 nm)/Au(300 nm) for the transmission lines. The coplanar waveguide has a center-line to side-line distance of 60 μm. The waveguides have a 60 μm-wide center-line, that tapers down to a 5.5 μm spaced 4 μm-wide center-line, as depicted in Figure 2. The magnetic load is a 20 μm x 4 μm strip partially covered by the

Au transmission lines, so current only flows through the magnetic stack only in the uncovered 5 x 4 μm opening.

The Hall bars where patterned via similar lithography process, as published elsewhere[57].

### Generation of picosecond-duration electrical pulses

In order to generate the picosecond pulses (schematic shown in Figure 2) we contact the left side of the transmission lines with a CPW 40GHz GBB® probe tip. We also contact the right side with another CPW 40GHz GBB® tip, and add a 50Ω resistor to close the circuit. We apply a constant voltage bias ($\Delta V$) through the left tip between -50V and +50V via a Keithley 2400 voltage source, while reading the average current. If no laser irradiation is incident, we can measure a dark (i.e. leakage) current due to the finite switch resistance (>10 MΩ). We then irradiate the photoswitch with either 1.5mW (0.3 μJ per pulse) from our 5kHz amplified laser or 30mW (0.37 nJ per pulse) from the 80MHz oscillator system. The pump is focused by a 15 cm lens to a (FWHM) radius of about 150 μm. When the switch is irradiated a photocurrent is generated, which we optimize by finely tuning the pump mirror. We note that photoswitch excitation with the high voltages and high pulse energies used for the switching experiments can result in longer electrical pulses durations (6 ps for switching experiments vs 3.7 ps for time-resolved experiments)[35].

### MOKE micrographs

The images where obtained with a home-made magneto-optical Kerr effect microscope. We use a 633nm light source in the Köhler configuration, a long-working distance 50x objective and a 6fps monochrome CCD. Due to the highly reflective Au transmission lines that are next to the magnetic section, we are very susceptible to pixel overflowing, which means the exposure time has to be kept short, which results in an important degradation of magneto-optical signal. As a solution we use 8x binning, without loss of spatial resolution, to speed up the camera up to 15fps and average 60 images to reduce noise and boost back the signal to noise ratio. Finally, mechanical drift and vibrations importantly degrade the quality of the images, in particular when live background subtraction is used in order to the enhance the magneto-optical contrast.

### Measurement of time-resolved dynamics

All the presented small-current induced dynamics where measured with the 80MHz oscillator laser system focused through a 50x objective into a ~1μm sized spot. The experiments where performed with no out-of-plane field, since at small excitations the sample naturally relaxed back between pulses, as is typical with low-excitation optical pump-probe experiments. We determined the zero delay time, i.e. the arrival of the electrical pulse, by monitoring the time-domain thermoreflectance (TDTR) response (see supp. mat.).

### Measurement of picosecond pulse electric field

A pump beam is focused on the photoconductive switch to generate the pulse. The pump beam is chopped by an optical chopper at about 300 Hz. We place the Teraspike® probe on top of the transmission lines. A probe beam is focused at the tip of Teraspike® probe, exciting the free-standing photoconductive switch. A Stanford Research Systems 865A Lock-In amplifier is used to directly measure the change in current induced by the transient electric field at the tip of the probe. This change in current is only present when both the electric field and probe beams coincide in time. Therefore, changing the delay between optical pump and probes allows us to sample the transient electric field associated with the picosecond current pulses. The resulting trace is a convolution of the real electric field and the Teraspike®'s response. We note that we do not measure the electric pulse at the sample position.

## Estimation of switch capacitance and energy dissipation

The capacitance of an interdigitated electrode (IDE) capacitor $C_{IDE}$ is roughly $C_{IDE} = (N-1)\epsilon_r\epsilon_0 A/d$ where $N$ is the number of electrodes, $\epsilon_r$ is the effective relative permittivity due to LT-GaAs substrate and air (measured as 15), $\epsilon_0$ is the vacuum permittivity, $A$ is the surface area of an electrode and $d$ is the center-to-center distance between electrodes. We find a capacitance of about $\sim 10^{-14}$ F for our photoconductive switch. As a second verification, it is also well known that the RC time constant due to the capacitance of the photoswitch limits the pulse duration of the generated pulses[18], which means we can also set an upper bound for the capacitance given by RC<3.7ps (which is our smallest measured pulse duration). Here, the characteristic impedance ($Z_0$) of the line plays the role of the resistor. The CPW line impedance was designed[58] to be $Z_0 = 70\Omega$, which means the capacitance is at most around $5.3 \cdot 10^{-14}$ F, consistent with our initial estimation. We will consider this upper-bound value as the capacitance of our switch to calculate the upper-bound energy dissipation in our experiments. For the measured bias voltage at the switching threshold voltage of $\Delta V = 40$ V, the energy stored is $\frac{1}{2}C_{max}\Delta V^2 \sim 50$ pJ. In a worst case scenario, if we assume a full discharge of the switch, no radiative losses, no transmission loses and assume a perfect absorption at the magnet, then all of the energy stored in the switch capacitor would be dissipated at the load. For our magnetic load dimensions, this corresponds to an energy density of $\sim 150$ MJ/m$^3$. Because the energy dissipation for a Gaussian current pulse is $E = \int J(t)^2 \rho \cdot dt = 0.75 \cdot J^2 \rho \tau_p$, where $\rho$ (= $81\ \mu\Omega \cdot$ cm ) is the measured resistivity of the magnet, we estimate the maximum peak current density for switching with $\tau_p = 6$ ps pulses to be about $J_c \sim 6 \cdot 10^{12}$ A/m$^2$.

# Picosecond Spin Orbit Torque Switching

Kaushalya Jhuria[1], Julius Hohlfeld[1], Akshay Pattabi[2], Elodie Martin[1], Aldo Ygnacio Arriola Córdova[1,3], Xinping Shi[4], Roberto Lo Conte[2], Sebastien Petit-Watelot[1], Juan Carlos Rojas-Sanchez[1], Gregory Malinowski[1], Stéphane Mangin[1], Aristide Lemaître[5], Michel Hehn[1], Jeffrey Bokor[2,6], Richard B. Wilson[4], Jon Gorchon[1,*]

## Supplementary materials:

[1]Université de Lorraine, CNRS, IJL, F-54000 Nancy, France [2]Department of Electrical Engineering and Computer Sciences, University of California, Berkeley, CA 94720, USA [3]Universidad Nacional de Ingeniería, Avenida Túpac Amaru 210, Rímac, Lima, Perú [4]Department of Mechanical Engineering and Materials Science and Engineering Program, University of California, Riverside, CA 92521, USA [5]Centre de Nanosciences et de Nanotechnologies (C2N), CNRS, Université Paris Sud, Université Paris-Saclay, 91120 Palaiseau, France [6]Lawrence Berkeley National Laboratory, 1 Cyclotron Road, Berkeley, CA 94720, USA
*e-mail: jon.gorchon@univ-lorraine.fr

## 1. Dependence of quasi-static critical current density on in-plane field

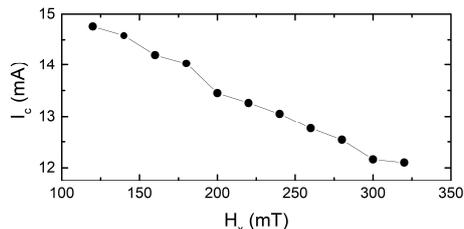

**Figure S1:** The critical current density for SOT switching with 100 μs pulses is inversely proportional to the in-plane $H_x$ field, as reported previously[23].

## 2. Estimation of resistivity and quasi-static critical current density

A four point resistivity measurement yielded an effective resistivity of $\rho = 81\ \mu\Omega\cdot cm$. Using previously measured values of resisitivity for 4-5nm Pt and Ta films deposited with the same sputtering system ($24$ and $200\ \mu\Omega\cdot cm$ respectively), and a parallel resistor model, we determine the correct effective resistivity by assuming that most of the current flows in the 13nm comprised by the 5nm Ta buffer, bottom 4nm Pt and top 4nm Ta layers, and neglecting the currents through the 1nm Co, Cu and Pt capping layers. We then estimate the critical current density by using again a parallel resistor model to calculate the amount of current going through Pt and Ta layers surrounding the Co/Cu bilayer.

## 3. Switched area vs current amplitude

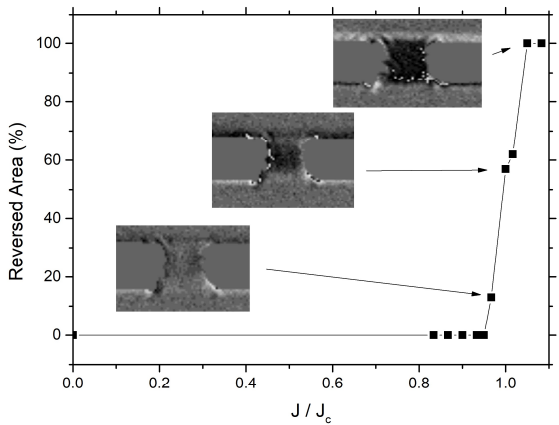

**Figure S2:** The switched area as a function the current density normalized by the critical current density ($J_c$). The reversal process is first happening where the current density is the highest, at mid-height, right in between the tips of the rounded gold electrodes. In order to fully switch the device a little more current density than $J_c$ is needed. We note that these experiments were performed on a different sample with a coplanar stripline design. Moreover, the magnetic section had a ~$5\mu m$ width, slightly wider than the main sample of the article (~$4\mu m$). In fact, in the main sample, these partial reversals were not clearly evidenced. We believe this could be due to the narrower section, or also due to the gold contacts being flatter, possibly resulting in a more homogeneous current distribution. We also note that in the experiments of Figure S2 the current pulse duration is unknown (not measured). We estimate it in between 3-10ps, from experience with similar devices.

## 4. Determination of zero-time delay from Time Domain Thermoreflectance

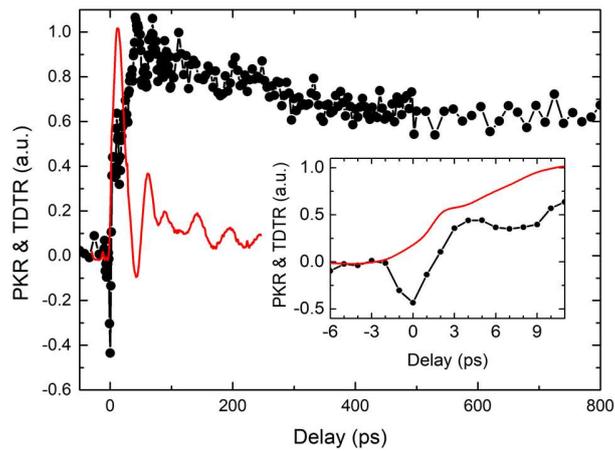

**Figure S3: Time-domain thermoreflectance (TDTR) in black and polar MOKE (PKR) response in red.** The TDTR allows us to set time-zero in our experiment. The electrons immediately respond to the heat pulse (negative peak at time-zero). The magnetic dynamics (red) equally start at the arrival of the pulse with no noticeable delay. Further work is needed to fully interpret the TDTR response.

## 5. Time-resolved switching dynamics

Time-resolved pump/probe measurements of the switching dynamics are not yet possible with our 80 MHz laser system, since the energy per pulse is not high enough to generate sufficiently intense current pulses. The time-resolved dynamics shown in was taken with almost the full pump power irradiating the switch, and the strongest in-plane field available in our setup. Probing the switching dynamics was also not possible with our 5 kHz amplified laser system. On our MOKE setup, measurements require a minimum probe power of ~60 μW to resolve Kerr rotations of ~100 μrad. The full amplitude of the hysteresis ($2M_s$) measured with the 80MHz oscillator system, shown in **Figure S4,** is ~485 μrad. Therefore, 60 μW would allow us to resolve the dynamics with a SNR of less than ~5:1 (without accounting for drift issues). However, at a 5 kHz rep rate, the per pulse energy for 60 μW causes damage to the sample when focused to sub-3 μm dimensions. Possible solutions could be to make the sample area bigger, and defocus the beam, or to use a higher repetition rate laser.

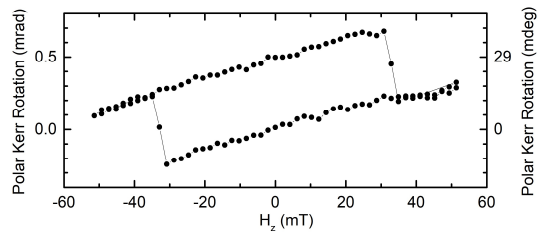

**Figure S4: Polar MOKE hysteresis on the CPW-embedded magnetic sample taken with the 80MHz laser system.**

## 6. HAMR-scenario and dependence of coercivity on number of pulses

In order to check for a heat assisted magnetic recording-like scenario, we injected single pulses at the switching threshold current, under no in-plane field, and monitored the variation of the coercive field (the applied out-of-plane field leading to ~50% reversal or more) with the number of applied current pulses. If the sample was to be heated near $T_c$, then any out-of-plane magnetic field would lead to reversal even after excitation with a single current pulse. However, we did not observe any switching when injecting a single pulse and applying an out-of-plane field as large as 93% of the switching field. In fact, we only observe a small decrease of ~30% in the coercivity when increasing the number of pulses by a factor of $10^5$ (see **Figure S5**). We conclude that the dissipation by the electrical pulse does not heat the Co film near $T_c$.

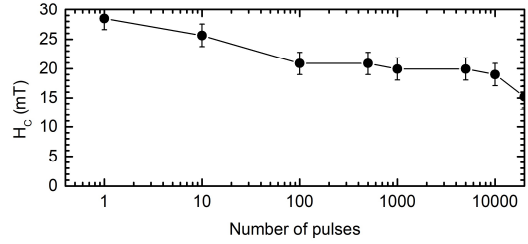

Figure S5: Coercivity as a function of the number of single 6ps pulses spaced every 200 μs.

## 7. Spatial dependence of magnetic dynamics

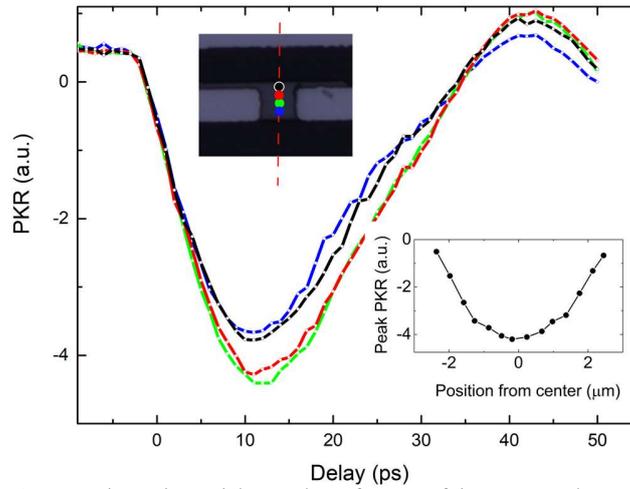

**Figure S6: Spatial dependence of dynamics.** Inset shows the peak (at 11ps) as a function of the y position (across the sample width). The signal drops as we get close to the edges because the probe no longer fully overlaps the magnet (the probe width is about 1.5 μm (FWHM), and the sample width is 4 μm. The dynamics are extremely similar across the surface of the sample. Experiments along the length of the magnet (x direction) also showed no major differences.

## 8. Macrospin model (table of parameters in section 7.4)

### 8.1. LLG dynamics

To model the ability of picosecond electrical pulses to reverse the magnetization via spin orbit torques, we solve the Landau-Lifshitz-Gilbert equation of motion with anti-damping like and field-like damping terms from the spin-current in the following form:

$$\frac{d\vec{M}}{dt} = -\gamma\mu_0\left(\vec{M}\times\vec{H}_{eff}\right) + \frac{\alpha}{M_s}\left(\vec{M}\times\frac{d\vec{M}}{dt}\right) - \theta_{SH}^{DL}\frac{C_s}{M_s}\left(\vec{M}\times\left(\vec{M}\times\vec{\sigma}\right)\right) + \theta_{SH}^{FL}C_s(\vec{M}\times\vec{\sigma}) \qquad \text{(eq.S1)}$$

where

$$C_s = \frac{\mu_B}{q_e}\frac{J_c}{d_0}\frac{1}{M_s} \qquad \text{(eq.S2)}$$

Here, $\gamma$ is the gyromagnetic ratio, $\mu_0$ is the vacuum permeability, $\alpha$ is the damping parameter, $M_s$ is the saturation magnetization, $\vec{\sigma}$ is the direction of spin-polarization, $\theta_{SH}^{DL}$ is the damping-like spin Hall angle, $\theta_{SH}^{FL}$ is the field-like spin Hall angle, $J_c$ is the current density, $d_0$ is the thickness of the magnetic layer, $q_e$ is the charge of the electron and $\mu_B$ is the Bohr magneton. The effective field in the first term consists of a magneto-crystalline anisotropy field, a demagnetization field, and any applied external field,

$$\vec{H}_{eff} = -\frac{1}{\mu_0 M_s} \frac{\partial F}{\partial \vec{M}} = \begin{bmatrix} H_x \\ H_y \\ H_z + \left(\frac{2K_z}{\mu_0 M_s} - M_s\right)m_z \end{bmatrix} \quad \text{(eq.S3)}$$

Here, $\vec{H}_x$, $\vec{H}_y$ and $\vec{H}_z$ are the x, y, and z-components of the external field, $K_z$ is the perpendicular anisotropy constant, and $\vec{M} = M_s[\vec{m}_x \quad \vec{m}_y \quad \vec{m}_z]^T$. The $-M_s\vec{m}_z$ term in the z-direction is the demagnetization field due to thin-film shape anisotropy. In this work, we use the macrospin approximation, i.e. we assume that the properties in Eq. (eq.S1) are independent of position. To solve for the dynamics, we first set $d\vec{M}/dt = 0$ and solve for the equilibrium orientation of the moment, $\vec{M}(t<0) = \vec{M}_0$. We identify the stable solution to $d\vec{M}/dt = 0$ by choosing the solution with the lowest free energy. We then use a finite-difference scheme to evolve $\vec{M}(t)$ forwards in time in response to a charge current $J_c(t)$. We evolve the magnetization forward in time with time-increments of $\Delta t = 1$ fs (we verified that the results were unchanged with smaller time-increments).

## 8.2. Temperature Dynamics

Ultrafast heating of a magnetic material leads to temperature-induced changes to the magnetic moment and the interfacial anisotropy. We estimate the temperature response of the metal film in our experiments by solving the heat-diffusion equation,

$$C\frac{dT}{dt} = \Lambda \frac{d^2T}{dx^2} + q(t) \quad \text{(eq.S4)}$$

Here, $T$ is the temperature, $C$ is the heat-capacity per unit volume, and $q(t)$ is the volumetric heating from either a laser or electrical pulse. Based on literature values of the heat-capacity of metals [1] and the thickness of each layer, we estimate an average value for $C$ of the multilayer of 2.6 J m$^{-3}$ K$^{-1}$ for our stack at room temperature. We fix $\Lambda$ according to the Wiedemann-Franz Law $\Lambda = L_0 T/\rho_e \sim 9$ Wm$^{-1}$K$^{-1}$, where $\rho_e = 81$ $\mu\Omega\cdot$cm is the measured electrical resistivity of the film. For the optical experiments described in section 7.3., we assume $q(t) = P_{abs}(t)/(\pi w_0^2 d)$, where $P_{abs}(t)$ is the absorbed laser power vs. time, $w_0$ is the 1/e$^2$ radius, and $d$ is the total film thickness of 16 nm. For electrical experiments, we set $q(t) = \rho_e J^2(t)$, where $J(t)$ is the charge current density. Solving Eq. (eq.S4) for $T(t)$ requires boundary conditions. We assume an adiabatic boundary condition at the metal film surface. We assume the heat-current $J_Q$ at the bottom of the metal film is limited by the interfacial thermal conductance $G_{int}$ between the Ta and sapphire,

$$J_Q = G_{int}T(z=d) \quad \text{(eq.S5)}$$

Typical values for the conductance between metal films and oxide substrates are 100-300 MWm$^{-2}$K$^{-1}$ [2-4]. We treat $G_{int}$ as a fit parameter and deduce $G_{int} \sim 170$ MWm$^{-2}$K$^{-1}$.

By using Eq. (eq.S4) to model the temperature response of the stack to heating, we are assuming that electrons, phonons, and spins are in thermal equilibrium with one another. Such an assumption is not always valid on picosecond time-scales, and nonequilibrium between thermal resevoirs can drive ultrafast magnetic phenomena [5]. In our experiments, nonequilibrium effects should be small due to the 4-6 picosecond pulse duration of the electrical experiments, together with the strong thermal coupling between electrons and phonons in the Co layer [6]. The picosecond time-scale for heating is much greater than the electron-phonon relaxation time in transition metals [7]. Therefore, we can estimate the nonequilibrium between electrons and phonons by assuming a quasi-steady-state condition where rate of heat absorption of electrons equals the rate of heat-loss to the phonons. In other words, we assume $q(t) \sim g_{ep}\Delta T_{ep}(t)$, where $g_{ep}$ is the electron-phonon volumetric energy transfer coefficient [8], which we take from Ref. [6]. For our estimated maximum electrical current density of $6\cdot10^{12}$ A m$^{-2}$, we can estimate the upper bound for the nonequilibrium during our experiments to be $\Delta T_{ep}(t) \sim 15$K, or about 25% of the average peak temperature rise at those currents.

## 8.3. Anisotropy Torques and Precessional Dynamics Caused by Ultrafast Heating

The anisotropy field and magnetization are both temperature dependent [9]. As a result, picosecond changes in the temperature of the Co film induce precessional dynamics [10], even in the absence of spin-orbit-torques or Oersted fields. To experimentally investigate the effect of temperature, we perform time-resolved magneto-optic Kerr effect measurements on the sample in the presence of an in-plane magnetic field. Prior to these time-resolved measurements, we orient the magnetic moment of the Cobalt with an out-of-plane magnetic field of 0.3 Tesla. After removing the out-of-plane magnetic field, we optical heat the sample surface with 250 fs duration pump pulses at a fluence of 0.7 J m$^{-2}$ under the application of an in-plane magnetic field. The transient temperature response causes precessional dynamics. We track the resulting out-of-plane component of the magnetic moment by monitoring the polar Kerr angle with a time-delayed probe pulse. We repeat this experiment with varied in-plane magnetic fields.

Fig. S7 shows the results of time-resolved magneto-optic Kerr measurements with an in-plane field of 0, 0.15, and 0.3 T on an identical magnetic stack which was co-grown on a glass substrate (due to the glass substrate, the interfacial conductance is lowered to $G_{int} \sim 100$ MWm$^{-2}$K$^{-1}$). The lines in Fig. S7 are best-fit model predictions for the data based on the LLG equations, including the thermal model of section 7.2. We describe those predictions in more detail below.

Using an optical multilayer calculation like described in Ref. [11], we estimate the absorbed fluence per pump pulse ($F$) is 0.3 J m-2. We estimate that this absorbed fluence should cause a per-pulse temperature rise of approximatively $F/(d \cdot C) \sim 7$ K. Comparing this temperature rise to our experimental measurements of the resulting $\Delta M_z$, allows us to quantify $dM_z/dT$ close to room temperature (i.e. for small heatings, just like in the SOT time-resolved dynamics). The amplitude of precession of $M_z(t)$ following heating provides information about $dK_z/dT$. The frequency and decay rate of precession allow us to determine the total anisotropy field and effective damping.

To theoretically quantify the effects of the temperature-evolution after optical heating we added temperature effects to Eq. (eq.S1-S3). We allow $M_s$ and $K_z$ in Eq. (eq.S1-S3) to evolve in time based on the predictions of our thermal model described in Eq. (eq.S4-S5). We follow Ref.[9], and assume the temperature dependencies of the magnetization and magneto-crystalline anisotropy to be described by

$$M_s(T) = M_s(0)[1 - (T/T_c)^{1.7}] \qquad \text{(eq.S6)}$$

$$K_z(T) = K_z(0)[M_s(T)/M_s(0)]^3 \qquad \text{(eq.S7)}$$

Here, $T_c$ is the Curie temperature, $M_s(0)$ is the magnetization at absolute zero, and $K_z(0)$ is the anisotropy constant at absolute zero.. We fix $M_s(T = 300 \, K)$ to $10^6$ A/m based on VSM measurements. We treat $K_z(T = 300 \, K)$ as a fit-parameter. The precessional frequency in Fig. eq.S2 depends only on the magnetic moment and total anisotropy field. With $M_s(300 \, K)$ fixed from VSM measurements, the only unknown parameter that affects the precessional frequency is $K_z(300 \, K)$. The best-fit value for $K_z(300 \, K)$ is $10^6$ Jm$^{-3}$, corresponding to an out-of-plane anisotropy field $B_K(300 \, K) = 2K_z/M_S - \mu_0 M_s$ of $\sim 0.8$ T. The fitting is only carried for data after several picoseconds, since the simple thermal model we are using (to avoid including too many parameters) does not account for the absorption profile of the laser pulse and assumes an instantaneous homogeneous heating of the magnetic stack, not capturing the initial thermal equilibration between layers. We fit for the effective value of $\alpha$ in eq.S1 by matching the model's prediction for the decay of oscillations to the data, and obtain a surprisingly large $\alpha \sim 0.23$. We note that the effective $\alpha$ includes effects such as inhomogeneous broadening [12]. The large damping could also explain why we could not detect the ferromagnetic resonance with our vector network analyzer. The only remaining model parameter is $T_c$. The value of $T_c$ determines $dM_z/dT$ and $dK_z/dT$ at room temperature. A best-fit to our data in Fig. S6 yields $T_c \sim 800$ K, in agreement with our extensive experience with similar samples grown in the same sputtering system. Then, eqs.S6-S7 predict $dM_z/dT \sim -10^3$ Am$^{-1}$K$^{-1}$ and $dK_z/dT \sim -4 \cdot 10^3$ Jm$^{-3}$K$^{-1}$, around room temperature.

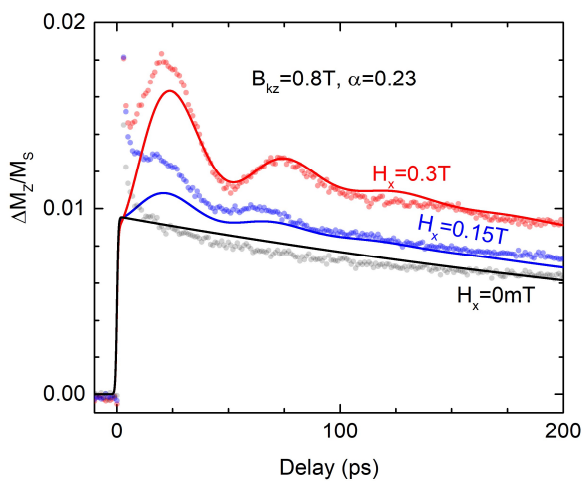

**Figure S7. Time-resolve magneto optic Kerr effect of temperature induced dynamics of the sample.** Markers are experimental data with 0 (black), 0.15 (blue) and 0.3 T (orange) in-plane applied magnetic fields. Lines are LLG macrospin fits.

## 8.4. Table of selected and fitted parameters

Below is the list of all used parameters during all the SOT simulations. We note that we fix most parameters using different sources, and only a few variable parameters are fitted (in red), with the benefit that each of them influences the dynamics differently. For example, the interface thermal conductance plays a big role in the recovery (long time-delay), whereas the field-like spin Hall angle is very important during the initial instants, while the current is on. Limitations of the model are discussed in the paper.

| | Name | Variable | Value | Units | Source |
|---|---|---|---|---|---|
| LLG | Saturation magnetization | $M_S$ (300K) | $10^6$ | Am⁻¹ | VSM measurement |
| | Magnetic anisotropy field | $B_K$ (300K) | 0.8 | T | Optical pump-probe data fit (Supplementary 7.3) |
| | Damping | $\alpha$ | 0.23 | --- | Optical data fit |
| | Thickness of magnetic layer | $d_0$ | 1 | nm | Estimated from growth rate calibrations |
| | Thickness of full metallic stack | $d$ | 16 | nm | Estimated from growth rate calibrations |
| | Damping-like spin Hall angle | $\theta_{SH}^{DL}$ | 0.3 | --- | Set as in similar stacks from article Ref. [13] |
| | Field-like spin Hall angle | $\theta_{SH}^{FL}$ | 0.05 | --- | Free parameter |
| | Current density | $J_c$ | variable | Am⁻² | Free parameter (values shown in Figure captions) |
| Thermal | Curie Temperature | $T_C$ | 800 | K | Optical data fit & experience from previous samples. |
| | Volumetric total heat capacity | $C$ | $2.6\cdot10^6$ | Jm⁻³K⁻¹ | Weighted average value from [1] Typical 1-3 MJm⁻³K⁻¹ |
| | Interface thermal conductance | $G$ | $170\cdot10^6$ | Wm⁻²K⁻¹ | Free parameter. Typical 100-300 MWm⁻²K⁻¹ [2-4]. |
| | Thermal conductivity | $\Lambda$ (300K) | 9 | Wm⁻¹K⁻¹ | Wiedemann-Franz Law $\Lambda = L_0 T/\rho_e \sim 10$ |
| | Electrical resistivity | $\rho_e$ | $81\cdot10^{-8}$ | Ωm | 4 point measurement |

**Table T1. Parameters of macrospin model**

## 8.5. Critical current density vs pulse duration

In this section we compute the final outcome of the magnetic moment of the Co layer after a single current pulse, as a function of current density and pulse duration ($\Delta t$). We increase the current density for a given pulse duration until the magnetization's final state (end of the pulse + 200ps) is reversed and record that current density as the critical current density. As shown in Figure S8, we compute two different scenarios: Only LLG (red circles) and LLG + thermal effects (black open circles). We remind that the results for strong currents (such as here, for switching) when including thermal effects, are not representative of our samples necessarily, since we do not know the exact temperature dependence of $M_S$ and $K_z$ away from room temperature. However, it is clear that ultrafast heating should lead to extra torques (just as in the case of optical pump-probe experiments) that should help the switching. Indeed, as Figure S8 shows, the critical current density (Fig S8a) and required energy (Fig S8b) is smaller when including thermal effects. The ratio between the required critical current densities (energies) in both models is plotted as a dotted orange line in FigS8a (b). With the used parameters, this ratio is at least ~1.5 thereby yielding at least a factor of 2 in energy gains.

When using the macrospin LLG model (red line in Figure S8), we find a critical current proportional to $\Delta t^{-1}$ (+ an offset), which could mean that the spin angular momentum for switching is constant below a certain pulse duration, as suggested in Ref.14. Surprisingly, also in Ref.14, a $\Delta t^{-2}$ relationship was found using similar macrospin simulations. When we include thermal effects (black empty circles) the relationship deviates from $\Delta t^{-1}$. However, we note that if we were to make a fit of the thermal estimates (black empty circles) using a $\Delta t^{-1}$ relationship within a reduced range (between ~100ps and ~1ns as in Ref.14) the fit would still work quite well (see blue dashed line) for that narrow range. This exercise highlights the difficulty in extracting the exact exponent when the range of the data is narrow.

Finally, it is interesting to look at the energy requirements on the right Figure S8b. Because of damping and anisotropy, as the current pulse gets wider than 10 ns, a large part of the angular momentum injected by the current pulses gets dissipated and thus even more current is required to keep increasing the precession angle, until switching is achieved. This results in an important increase in the required energy. On the other hand, when going into the very short pulse regime, even if less angular momentum is "wasted", more and more current density is required, resulting in stronger Joule heating. In between, around 10-20 ps for the used parameters (not far from the FMR half-period), a minimum of required energy will be found. Curiously it is around this minimum point that we also find the least difference between the two model's estimated critical energies.

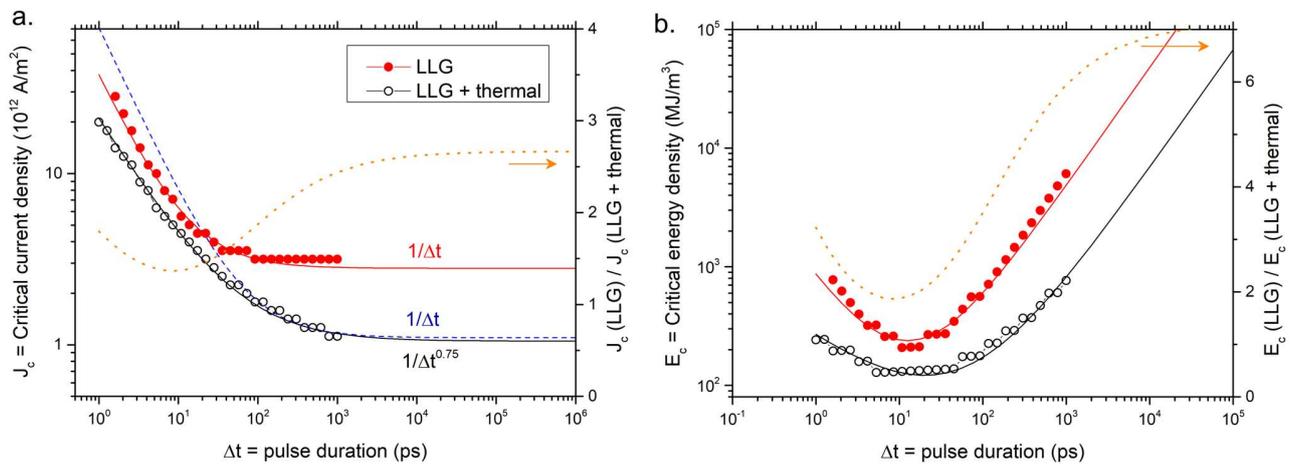

**Figure S8. Simulated critical current densities as a function of current pulse duration.** a. Critical current density and b. critical energy as a function of the pulse duration. Red filled circles are obtained using the LLG model and the black open circles are obtained when adding the thermal model. Lines are ~$1/\Delta t^{\beta} + J_{c0}$ fits, as discussed in the text, for comparison with previous work from Ref.14. The orange dot line in a. (b.) corresponds to the ratio of critical current densities (energies) between the LLG and the LLG + thermal model. In plane field is 0.16T and the rest of parameters are presented in table T1.

## 8.6. Effect of Heating on Ultrafast SOT Dynamics

To evaluate the role of temperature rises on the switching dynamics, we use the model parameters described above, to estimate the change in dynamics that results from the addition of temperature-induced precessional dynamics. We set an in-plane magnetic field of 0.16T and a 6ps wide pulse. The results of these simulations for two opposite current directions (plotted as blue and red) are shown in Figure S9. Using the full thermal model (Figure S9b) the critical current density is estimated to around $6 \cdot 10^{12}$ A/m$^2$ whereas when using a pure LLG with no temperature dependence (Figure S9a) we need to increase the current up to $9 \cdot 10^{12}$ A/m$^2$ in order to observe switching. This corresponds to about a twofold increase in energy dissipation/consumption. The speed of reversal can vary quite a bit depending on the current density, but we observe minimum switching times around 16ps (for the crossing of zero magnetization), for the shown parameters.

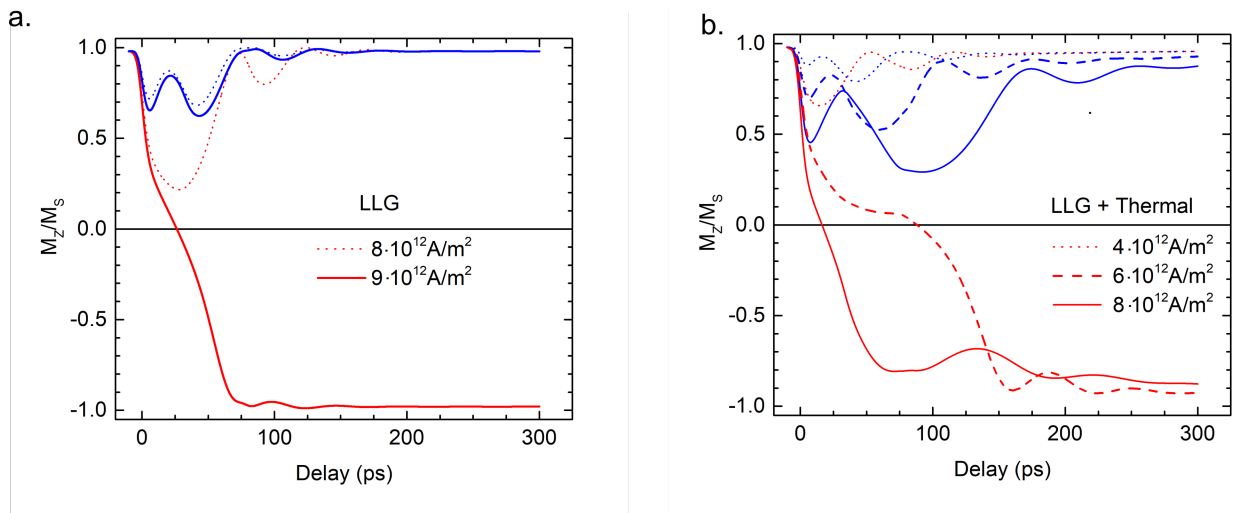

**Figure S9. Strong current dynamics predicted with a 6 ps pulse.** Dynamics of the normalized out-of-plane magnetization Mz as obtained from numerical solutions of the LLG model that a) exclude and b) include the variations of material parameters with temperature. All blue curves correspond to a positive current flow, and red curves to a negative current flow. The fastest reversal (crossing of zero) takes about 16ps in our simulations, but only reaches 80% of saturation after 50ps. We assume an in-plane field of about 0.16T. The rest of parameters are in table T1.

Interestingly, even if we set the spin Hall angles to zero (and thus have zero SOT), if we increase the current density to have enough of a temperature rise, then a strong enough thermal anisotropy torque will be present and lead to switching. We show this in the simulation of Figure S10. This was also experienced in YIG by using ultrafast optical pulses [15], and the important requirementss are: a) having dissimilar thermal derivatives of M and K, and b) to perform the experiment under a constant field perpendicular to the magnetic anisotropy axis.

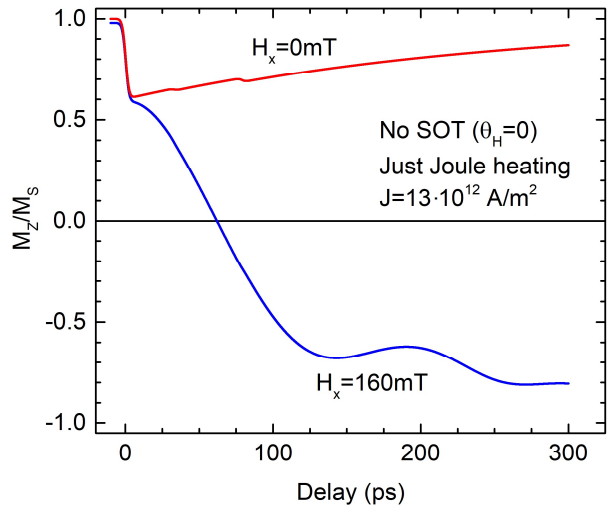

**Figure S10. Switching with a 6 ps pulse purely via the thermal anisotropy torque.** We set the spin Hall angle to zero, and simulate the dynamics of the out-of-plane magnetization under the influence of Joule heating due to a 6 ps pulse. In case an in-plane 160mT field is applied (blue line) we observe switching, but if no in-plane is applied (red line), switching is no longer possible.

References:

1. Y. S. Touloukian (Editor), Recommended Values of the Thermophysical Properties of Eight Alloys, Major Constituents and their Oxides Thermophysical Properties Research Center, Purdue University, Lafayette, Indiana, February, 1966

2. Hopkins, P.E., Thermal transport across solid interfaces with nanoscale imperfections: Effects of roughness, disorder, dislocations, and bonding on thermal boundary conductance. ISRN Mechanical Engineering, 2013. 2013.

3. Wilson, R., B.A. Apgar, W.-P. Hsieh, L.W. Martin, and D.G. Cahill, Thermal conductance of strongly bonded metal-oxide interfaces. Physical Review B, 2015. 91(11): p. 115414.

4. Monachon, C., L. Weber, and C. Dames, Thermal boundary conductance: A materials science perspective. Annual Review of Materials Research, 2016. 46: p. 433-463.

5. Beaurepaire, E., J.-C. Merle, A. Daunois, and J.-Y. Bigot, Ultrafast spin dynamics in ferromagnetic nickel. Physical review letters, 1996. 76(22): p. 4250.

6. Verstraete, M.J., Ab initio calculation of spin-dependent electron–phonon coupling in iron and cobalt. Journal of Physics: Condensed Matter, 2013. 25(13): p. 136001.

7. Allen, P.B., Empirical electron-phonon λ values from resistivity of cubic metallic elements. Physical Review B, 1987. 36(5): p. 2920.

8. Wilson, R.B., Y. Yang, J. Gorchon, C.-H. Lambert, S. Salahuddin, and J. Bokor, Electric current induced ultrafast demagnetization. Physical Review B, 2017. 96(4): p. 045105.

9. Lee, K.-M., J.W. Choi, J. Sok, and B.-C. Min, Temperature dependence of the interfacial magnetic anisotropy in W/CoFeB/MgO. AIP Advances, 2017. 7(6): p. 065107.

10. Atxitia, U., O. Chubykalo-Fesenko, N. Kazantseva, D. Hinzke, U. Nowak, and R.W. Chantrell, Micromagnetic modeling of laser-induced magnetization dynamics using the Landau-Lifshitz-Bloch equation. Applied physics letters, 2007. 91(23): p. 232507.

11. Yang, Y., R.B. Wilson, J. Gorchon, C.-H. Lambert, S. Salahuddin, and J. Bokor, Ultrafast magnetization reversal by picosecond electrical pulses. Science Advances, 2017. 3(11).

12. Lattery, D.M., J. Zhu, D. Zhang, J.-P. Wang, P.A. Crowell, and X. Wang, Quantitative analysis and optimization of magnetization precession initiated by ultrafast optical pulses. Applied Physics Letters, 2018. 113(16): p. 162405.

13. Woo, S., Mann, M., Tan, A. J., Caretta, L. & Beach, G. S. D. Enhanced spin-orbit torques in Pt/Co/Ta heterostructures. Appl. Phys. Lett. 105, 212404 (2014)

14. Garello, K. et al. Ultrafast magnetization switching by spin-orbit torques. Appl. Phys. Lett. 105, 1–12 (2014)

15. Davies, C. S. et al. Anomalously Damped Heat-Assisted Route for Precessional Magnetization Reversal in an Iron Garnet. Phys. Rev. Lett. 122, 27202 (2019)